\documentclass{llncs}

\input epsf

\begin{document}

\title{OCB: A Generic Benchmark to Evaluate
       the Performances of Object-Oriented Database Systems}

\author{J\'{e}r\^{o}me Darmont\\Bertrand Petit\\Michel Schneider}

\institute{Laboratoire d'Informatique (LIMOS)\\
           Universit\'{e} Blaise Pascal -- Clermont-Ferrand~II\\
           Complexe Scientifique des C\'{e}zeaux\\
           63177 Aubi\`{e}re Cedex\\
           FRANCE\\
           {\em darmont@libd1.univ-bpclermont.fr}}

\maketitle

\begin{abstract}

We present in this paper a generic object-oriented benchmark (the Object Clustering
Benchmark) that has been designed to evaluate the performances of clustering policies in
object-oriented databases. OCB is generic because its sample database may be customized to
fit the databases introduced by the main existing benchmarks (e.g., OO1). OCB's current form
is clustering-oriented because of its clustering-oriented workload, but it can be easily adapted
to other purposes. Lastly, OCB's code is compact and easily portable. OCB has been implemented
in a real system (Texas, running on a Sun workstation), in order to test a specific
clustering policy called DSTC. A few results concerning this test are presented.


\end{abstract}

\section{Introduction}

        This study originates from the design of clustering algorithms to improve the performance
of object-oriented databases. The principle of clustering is to store related objects close
together on secondary storage, so that when an object is accessed from disk, all its related objects
are also loaded into the main memory. Subsequent accesses to these related objects are
thus main memory accesses, instead of much slower I/Os.

        But clustering involves some overhead (to gather and maintain usage statistics, to reorganize
the database...), so it is not easy to determine the real impact of a given clustering heuristic
on the overall performances. Hence, clustering algorithms are validated only if performance
tests demonstrate their actual value.

        The validation of clustering methods can be achieved by several ways. First, mathematical
analysis can be used to ascertain the complexity of a clustering algorithm \cite{CHABRIDON93}.
Although mathematical analysis provides exact results, it is very difficult to take
into account all the parameters defining a real system. Hence, simplification hypothesis are
made, and results tend to differ from reality. Simulation may also be used, and offers several
advantages \cite{DARMONT97}. First, clustering algorithms that are possibly implemented on
different OODBs and/or operating systems can be compared within the same environment,
and thus on the same basis. A given algorithm can also be tested on different platforms to determine
how its behavior might be influenced by its host system. Simulation also allows the {\em a
priori} modeling of research prototypes before they are actually implemented in an OODB.
Eventually, the most customary mean to measure the performances of DBMSs in general is
the use of benchmarks, that directly gauge the response of an existing system, and, {\em a fortiori},
the performances of a clustering algorithm implemented in an existing system. However, the
usual general purpose benchmarks are not well suited to the evaluation of clustering algorithms,
that are very data dependent.

        Some authors propose dedicated tools to evaluate the performances of their own clustering
heuristic. We preferred to design generic tools, in order to be able to compare different
algorithms on the same basis, using standard and easy to implement metrics. It is actually interesting
to compare clustering policies together, instead of comparing them to a non-clustering
policy. We can also use different platforms to test a given algorithm. We are actually
involved in the development of both simulation models and a benchmark. We focus in
this paper on the latter, a generic, clustering-oriented benchmark called the Object Clustering
Benchmark (OCB).

        The remainder of this paper is organized as follows. Section~2 presents the most popular
benchmarks for evaluating the performances of OODBs. Our own benchmark, OCB, is then
described in Section~3. Section~4 presents experiments we performed to validate our benchmark.
Section~5 eventually concludes this paper and provides future research directions.

\section{Related Work}

        Benchmarking the performances of an OODB consists of performing a set of tests in order
to measure the system response under certain conditions. Benchmarks are used to compare
the global performances of OODBs, but also to illustrate the advantages of one system or another
in a given situation, or to determine an optimal hardware configuration (memory buffer
size, number of disks...) for a given OODB and/or application.
Several well-known standard object-oriented benchmarks are used nowadays. The presentation of three of them (OO1,
HyperModel, and OO7) follows.

        Typically, a benchmark is constituted of two main elements:
\begin{itemize}
\item a database (a conceptual schema, and a database generation method);
\item a workload (a set of operations to perform on the database, e.g., different kind of queries;
and a protocol detailing the execution of these operations).
\end{itemize}

\subsection{OO1}

        OO1 (Objects Operations 1), sometimes called the "Cattell Benchmark" \cite{CATTELL91},
is customarily used to evaluate the performances of both relational and object-oriented
DBMSs.

\subsubsection{OO1 Database:}

%        OO1's database (Fig.~\ref{fig-1}) is based on two classes: {\em Part} and {\em Connection}. The parts are
OO1's database is based on two classes: {\em Part}, and {\em Connection}. The parts are
composite elements that are connected (through {\em Connection} objects) to three other parts. Each
connection references two parts: the source ({\em From}), and the destination part ({\em To}).

%\begin{figure}[htb]
%\epsfxsize=18.5cm
%\vspace*{-0.85cm}
%\centerline{\epsffile{fig-1.eps}}
%\vspace*{-20cm}
%\caption{OO1 database schema (UML Static Structure Diagram)}
%\label{fig-1}
%\end{figure}
%\footnote{Unified Modeling Language}

        The database is generated the following way:
\begin{enumerate}
\item create all the {\em Part} objects and store them into a dictionary;
\item for each part, randomly choose three other parts and create the associated connections.
\end{enumerate}

        The locality of reference (objects are often linked to relatively close objects) is simulated
by a reference zone. I.e., {\em Part} \#i is randomly linked to parts which {\em Id} are in the interval
[{\em Id}-{\em RefZone}, {\em Id}+{\em RefZone}]. The probability that the links are determined this way is 0.9. Otherwise,
the linked parts are chosen totally at random.

\subsubsection{OO1 Workload:}

        OO1 performs three types of operations. Each of them is run 10 times. Response time is
measured for each run.

\begin{itemize}
\item {\em Lookup:} access to 1000 randomly selected parts.
\item {\em Traversal:} randomly select a root part, then explore the corresponding part tree (in depth
first) through the {\em Connect} and {\em To} references, up to seven hops (total of 3280 parts, with
possible duplicates). Also perform a {\em reverse traversal} by swapping the {\em To} and {\em From} directions.
\item {\em Insert:} add 100 parts, and the corresponding connections, to the database. Commit the
changes.
\end{itemize}

\subsubsection{Comments:}

        OO1 is a simple benchmark, and hence is very easy to implement. It was used to test a
broad range of systems, including object-oriented DBMSs, relational DBMSs, and other systems
like Sun's INDEX (B-tree based) system. Its visibility and simplicity actually make of
OO1 a standard for OODB benchmarking. However, its data model is too elementary to
measure the elaborate traversals that are common in many types of applications, like engineering
applications. Furthermore, OO1 only supports simple navigational and update tasks,
and has no notion of complex objects (e.g., composite objects).

\subsection{The HyperModel Benchmark}

        The HyperModel Benchmark (also called the Tektronix Benchmark in the literature)
\cite{ANDERSON90}\cite{BERRE91} offers a more complex database than OO1. Furthermore, it is
recognized for the richness of the tests it proposes.

\subsubsection{HyperModel Database:}

        The HyperModel Benchmark is based on an extended hypertext model. Hypertext is a
%generic graph structure consisting of nodes and links (Fig.~\ref{fig-2}). The main characteristic of
generic graph structure consisting of nodes and links. The main characteristic of
this database is the various relationships existing between classes: {\em inheritance} (the attributes
of a {\em Node} object may be inherited from another {\em Node} object), {\em aggregation} (an instance of the
{\em Node} class may be composed of one or several other instances), and eventually {\em association}
(two {\em Node} objects may be bound by an oriented link).

%\begin{figure}[htb]
%\epsfxsize=18.5cm
%\vspace*{-0.85cm}
%\centerline{\epsffile{fig-2.eps}}
%\vspace*{-15.75cm}
%\caption{HyperModel conceptual schema (UML Static Structure Diagram)}
%\label{fig-2}
%\end{figure}

\subsubsection{HyperModel Workload:}

        The benchmark consists of 20 operations. To measure the time to perform each operation,
the following sequence is followed.
\begin{enumerate}
\item {\em Setup:} prepare 50 inputs to the operations (the setup is not timed);
\item {\em Cold run:} run the operation 50 times, on the 50 inputs precomputed in the setup phase;
then, if the operation is an update, commit the changes once for all 50 operations;
\item {\em Warm run:} repeat the operation 50 times with the same input to test the effect of caching;
again, perform a commit if the operation was an update.
\end{enumerate}

        The 20 possible operations belong to seven different kinds:
\begin{itemize}
\item {\em Name Lookup:} retrieve one randomly selected node;
\item {\em Range Lookup:} retrieve the nodes satisfying a range predicate based on an attribute value;
\item {\em Group Lookup:} follow the relationships one level from a randomly selected starting node;
\item {\em Reference Lookup:} reverse Group Lookup;
\item {\em Sequential Scan:} visit all the nodes;
\item {\em Closure Traversal:} Group Lookup up to a predefined depth;
\item {\em Editing:} update one node.
\end{itemize}

\subsubsection{Comments:}

        The HyperModel Benchmark possesses both a richer schema, and a wider extent of operations
than OO1. This renders HyperModel potentially better than OO1 to measure the performances
of engineering databases. However, this added complexity also makes HyperModel
harder to implement, especially since its specifications are not as complete as OO1's. Lastly,
the HyperModel Benchmark still has no notion of complex object.

\subsection{OO7}

        OO7 \cite{CAREY94} is a more recent benchmark than OO1 and HyperModel, and hence
uses the structures described in the previous paragraphs to propose a more complete benchmark,
and to simulate various transactions running on a diversified database.

\subsubsection{OO7 Database:}

        OO7's database is based on a conceptual model that is very close to the HyperModel
%Benchmark's, though it contains a higher number of classes (Fig.~\ref{fig-3}). Four kinds of links are
Benchmark's, though it contains a higher number of classes. Four kinds of links are
also supported (IS-A, 1-1, 1-N, M-N). There are three sizes of the OO7 database: small, medium,
and large.

%\begin{figure}[htb]
%\epsfxsize=15cm
%\vspace*{-0.8cm}
%\centerline{\epsffile{fig-3.eps}}
%\vspace*{-8.25cm}
%\caption{OO7 conceptual schema (UML Static Structure Diagram)}
%\label{fig-3}
%\end{figure}

\subsubsection{OO7 Workload:}

        The range of transactions offered by OO7 is also close to HyperModel's. Three main
groups may be identified:
\begin{itemize}
\item {\em Traversals} browse the object graph using certain criteria. These traversals are very close to
OO1's. There are ten different operations that apply depending on the database characteristics
(basically, its size);
\item {\em Queries} retrieve objects chosen in function of various criteria. There are eight kinds of queries;
\item {\em Structural Modification Operations} deal with object insertion and deletion (two operations).
\end{itemize}

\subsubsection{Comments:}

        OO7 attempts to correct the flaws of OO1, and HyperModel. This is achieved with a rich
schema and a comprehensive set of operations. However, if OO7 is a good benchmark for
engineering applications (like CAD, CAM, or CASE), it might not be the case for other types
of applications based on objects. Since its schema is static, it cannot be adapted to other purposes.
Eventually, OO7 database structure and operations are nontrivial, hence making the
benchmark difficult to understand, adapt, or even implement (yet, to be fair, OO7 implementations
are available by anonymous FTP).

\section{The Object Clustering Benchmark}

        OCB's first purpose is to test the performances of clustering algorithms within object-oriented
systems. Hence, it is structured around a rich object base including many different
classes (and thus many different object sizes, numbers of references, etc.), and numerous types
of references (allowing the design of multiple interleaved hierarchies). OCB's workload is
purposely clustering-oriented, but can be easily extended to be fully generic as well.

        OCB's flexibility is also achieved through an extensive set of parameters that allow the
benchmark to be very adaptive. Many different kinds of object bases can be modeled with
OCB, as well as many different kinds of applications running on these databases. Since there
exists no canonical OODB application, this is an important feature. Eventually, the great majority
of these parameters are very easy to settle.

\subsection{OCB Justification}

        We initially felt the need for a clustering-oriented benchmark because the existing
benchmarks are not adapted to test the performances of most kinds of clustering algorithms,
including semantic clustering algorithms. General purpose benchmarks are useful when testing
the performances of an OODBMS as a whole, but inherently do not model any specific
application, even if most existing benchmarks were designed in a CAD/CAM/CASE context.
Hence, they are not well suited to benchmark the performances of clustering algorithms. Some
of their queries simply cannot benefit from any clustering, e.g., queries that scan through the
whole database, or random accesses. Furthermore, clustering is very data dependent, what is
not taken into account by the synthetic benchmarks, that all adopt a somewhat simple database
schema. Consequently, we designed a rich and complex database (though very easy to code
and generate), and a set of adapted queries.

        OCB's main characteristic is indeed its double aptitude to be both generic and clustering-oriented.
The clustering orientation definitely comes from the workload, but the set of transactions
we use can be easily extended to achieve full genericity. On the other hand, OCB's
database is wholly generic. OCB can be easily either parameterized to model a generic application,
or dedicated to a given type of object base and/or application.

        The last version of OCB (currently in development) also supports multiple users, in a
very simple way (using processes), which is almost unique. As far as we know, only OO7 has
a multi-user version also in development. OO1 was designed as multi-user, but the published
results only involve one single user.

        Eventually, OCB's code is very compact, and easy to implement on any platform. OCB is
currently implemented to benchmark the Texas persistent storage system for C++, coupled
with the DSTC clustering technique. The C++ code is less than 1,500 lines long. OCB is also
being ported into a simulation model designed with the QNAP2 simulation software, that supports
a non object-oriented language close to Pascal. The QNAP2 code dealing with OCB is
likely to be shorter than 1,000 lines.

\subsection{OCB Database}

\begin{figure}[htb]
\epsfxsize=15cm
\vspace*{-0.85cm}
\centerline{\epsffile{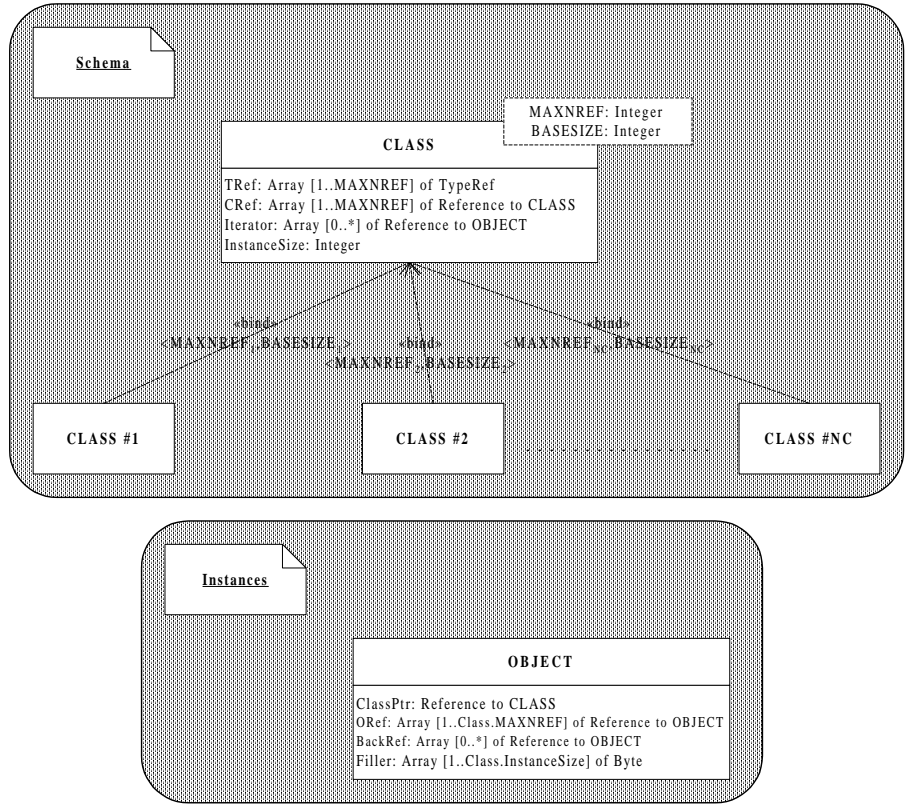}}
\vspace*{-9cm}
\caption{OCB database schema (UML Static Structure Diagram)}
\label{fig-4}
\end{figure}

        OCB's database is both rich and simple to achieve, very tunable, and thus highly generic.
The database is constituted of a predefined number of classes ({\em NC}), all derived from the same
metaclass (Fig.~\ref{fig-4}). A class is defined by two parameters: {\em MAXNREF}, the maximum number
of references present in the class' instances; and {\em BASESIZE}, the class' basic size (increment
size used to compute the {\em InstanceSize} after the inheritance graph is processed during the database
generation). Note that, on Fig.~\ref{fig-4}, the Unified Modeling Language (UML) "bind" clause indicates that classes are
instanciated from the metaclass using the parameters between brackets. Since different references
can point to the same class, 0-N, 1-N, and M-N links are implicitly modeled. Each reference
has a type. There are {\em NREFT} different types of reference. A reference type can be, for
instance, a type of inheritance, aggregation, user association, etc. After instanciation of the
database schema, an object points to at most {\em MAXNREF} objects from the iterator of the class
referenced by this object's class.

        The database generation proceeds through three chief steps.
\begin{enumerate}
\item Instanciation of the {\em CLASS} metaclass into {\em NC} classes: creation of the classes without any
reference, then selection of the classes referenced by each class. The type of the references
({\em TRef}) can be either randomly chosen according to the {\em DIST1} random distribution, or fixed
{\em a priori}. The referenced classes belong to an [{\em INFCLASS}, {\em SUPCLASS}] interval that models
a certain locality of reference at the class level. The class reference selection can be either
performed at random according to the {\em DIST2} random distribution, or set up {\em a priori}. NIL
references are possible.
\item Check-up of the database consistency: suppression of all the cycles and discrepancies
within the graphs that do not allow them (e.g., the inheritance graph, or composition hierarchies).
\item Instanciation of the {\em NC} classes into {\em NO} objects: creation of the objects, first without any
reference (their class is randomly determined according to the {\em DIST3} random distribution),
then random selection of the objects referenced by each object. The referenced objects belong
to an [{\em INFREF}, {\em SUPREF}] interval that models the locality of reference as introduced
by OO1. The object reference random selection is performed according to the {\em DIST4} random
distribution. Reverse references ({\em BackRef}) are instanciated at the same time the direct
links are.
\end{enumerate}

        The full database generation algorithm is provided in Fig.~\ref{fig-5}.

{\scriptsize
\begin{verbatim}
// Schema: Classes
For i = 1, NC do
  For j = 1, MAXNREF (i) do
    Class (i).TRef (j) = RAND (DIST1, 1, NREFT)
  End for
  Class (i).InstanceSize = BASESIZE (i)
End for

// Schema: Inter-classes references
For i = 1, NC do
  For j = 1, MAXNREF (i) do
    Class (i).CRef (j) = RAND (DIST2, INFCLASS, SUPCLASS)
  End for
End for

// Schema: Graph consistency for hierarchies without cycles
For i = 1, NC do
  For j = 1, MAXNREF (i) do
    If In_No_Cycle (Class (i).TRef (j)) then
      // Browse through class CRef (j) graph,
      // following the TRef (j) references
      If Class (i) belongs to the graph or a cycle is detected then
        Class (i).CRef (j) = NULL
      Else
        If Is_Inheritance (Class (i).TRef (j)) then
          // Browse through class CRef (j) inheritance graph
          // and add BASESIZE (i) to InstanceSize for each subclass
        End if
      End if
    End if
  End for
End for

// Instances: Objects
For i = 1, NO do
  Object (i).ClassPtr = RAND (DIST3, 1, NC)
  Object (i).ClassPtr.Add_Object_Into_Iterator (Object (i))
End for

// Instances: Inter-objects references
For i = 1, NC do
  For j = 1, Class (i).Get_Iterator_Count() do
    For k = 1, MAXNREF (i) do
      l = RAND (DIST4, INFREF, SUPREF)
      Iterator (i).Object (j).ORef (k) = Class (CRef(k)).Iterator (l)
      Add_BackRef (Class (CRef(k)).Iterator (l), Iterator (i).Object (j))
    End for
  End for
End for
\end{verbatim}}
\begin{figure}[hb]
\vspace*{-1.0cm}
\caption{OCB database generation algorithm}
\label{fig-5}
\end{figure}

{\em Note:} The random numbers used in the database creation are generated by the Lewis-Payne
random generator.

        The database parameters are summarized in Table~\ref{tab-1}.

\begin{table}[htb]
\begin{center}
\begin{tabular}{|c|c|c|}
\hline
{\bf Name} &
{\bf Parameter} &
{\bf Default value} \\
\hline
NC &
Number of classes in the database &
20 \\
MAXNREF (i) &
Maximum number of references, per class &
10 \\
BASESIZE (i) &
Instances base size, per class &
50 bytes \\
NO &
Total number of objects &
20000 \\
NREFT &
Number of reference types &
4 \\
INFCLASS &
Inferior bound, set of referenced classes &
1 \\
SUPCLASS &
Superior bound, set of referenced classes &
NC \\
INFREF &
Inferior bound, set of referenced objects &
1 \\
SUPREF &
Superior bound, set of referenced objects &
NO \\
DIST1 &
Reference types random distribution &
Uniform \\
DIST2 &
Class references random distribution &
Uniform \\
DIST3 &
Objects in classes random distribution &
Uniform \\
DIST4 &
Objects references random distribution &
Uniform \\
\hline
\end{tabular}
\vspace{5pt}
\caption{OCB database parameters}
\label{tab-1}
\end{center}
\end{table}

\subsection{OCB Workload}

        Since benchmarking the performances of clustering algorithms is our main goal, we focused
OCB's workload on a set of transactions that explore the effects of clustering. Hence,
we excluded at once some kinds of queries that simply cannot benefit from any clustering effort,
e.g., creation and update operations, or HyperModel Range Lookup and Sequential Scan
\cite{DARMONT97}.

\begin{figure}[htb]
\epsfxsize=15cm
\vspace*{-0.8cm}
\centerline{\epsffile{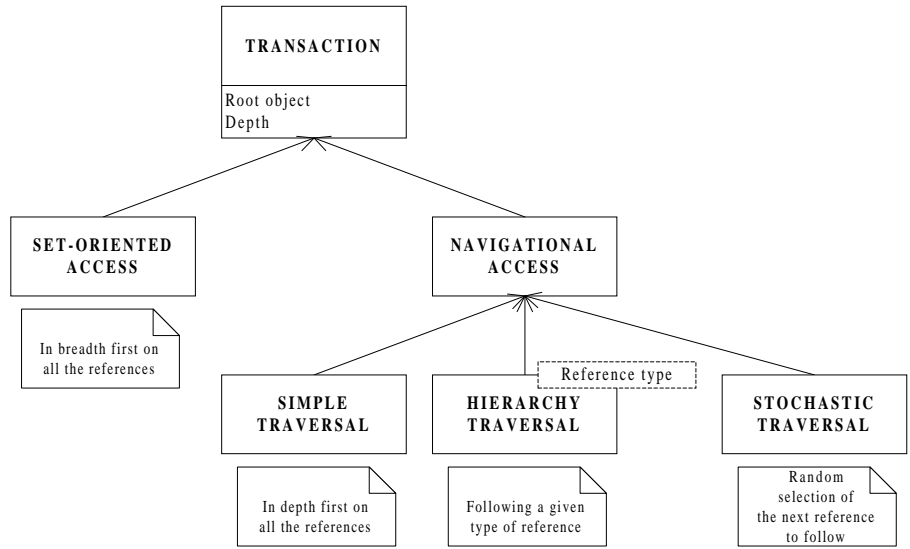}}
\vspace*{-12.75cm}
\caption{OCB transaction classes (UML Static Structure Diagram)}
\label{fig-6}
\end{figure}

        To model an appropriate workload for our needs, we used the types of transactions identified
by \cite{BULLAT96b} and \cite{MCIVER94} (Fig.~\ref{fig-6}). These operations are at the same time
well suited to explore the possibilities offered by clustering, and they can be tailored to model
different kinds of applications. They are basically divided into two types: set-oriented accesses,
and navigational accesses, that have been empirically found by \cite{MCIVER94} to match
breadth-first and depth-first traversals, respectively. Navigational accesses are further divided
into simple, depth first traversals, hierarchy traversals that always follow the same type of
reference, and finally stochastic traversals that select the next link to cross at random. Stochastic
traversals approach Markov chains, that simulate well the access patterns caused by
real queries, according to \cite{TSANGARIS92}. At each step, the probability to follow reference
number {\em N} is $p(N) = 1/2^N$.
Each type of transaction proceeds from a randomly chosen root
object (according to the {\em DIST5} random distribution), and up to a predefined depth depending
on the transaction type. All these transactions can be reversed to follow the links backwards
("ascending" the graphs).
%The C++ code for a hierarchy traversal is provided as an example in Fig.~\ref{fig-7}.

%{\scriptsize
%\begin{verbatim}
%void HierarchyTraversal (Object *Current, int CurrentDepth,
%                         int Depth, int ReferenceType) {

%  int i;

%  if (CurrentDepth<Depth) { // The traversal is not over
%    // All the references are considered
%    for (i=0; i<Current->ClassPtr->MAXNREF; i++) {
%      if ((Current->ORef[i]!=NULL)
%      && (Current->ClassPtr->TRef[i]==ReferenceType)) {
%        // Access to all non NULL references of type ReferenceType
%        HierarchyTraversal(Current->ORef[i], CurrentDepth+1,
%                           Depth, ReferenceType);
%      }}}
%};
%\end{verbatim}}
%\begin{figure}[hb]
%\vspace*{-1.0cm}
%\caption{Hierarchy traversal C++ code}
%\label{fig-7}
%\end{figure}

        The execution of the transactions by each client (the benchmark is to be multi-user) is
organized according to the following protocol:
\begin{enumerate}
\item cold run of {\em COLDN} transactions which types are determined randomly, according to predefined
probabilities. This step's purpose is to fill in the cache in order to observe the real
(stationary) behavior of the clustering algorithm implemented in the benchmarked system;
\item warm run of {\em HOTN} transactions.
\end{enumerate}

        A latency time {\em THINK} can be introduced between each transaction run.

        The OCB workload parameters are summarized in Table~\ref{tab-2}.

\begin{table}[htb]
\begin{center}
\begin{tabular}{|c|c|c|}
\hline
{\bf Name} &
{\bf Parameter} &
{\bf Default value} \\
\hline
SETDEPTH &
Set-oriented Access depth &
3 \\
SIMDEPTH &
Simple Traversal depth &
3 \\
HIEDEPTH &
Hierarchy Traversal depth &
5 \\
STODEPTH &
Stochastic Traversal depth &
50 \\
\hline
COLDN &
Number of transactions executed during cold run &
1000 \\
HOTN &
Number of transactions executed during warm run &
10000 \\
THINK &
Average latency time between transactions &
0 \\
PSET &
Set Access occurrence probability &
0.25 \\
PSIMPLE &
Simple Traversal occurrence probability &
0.25 \\
PHIER &
Hierarchy Traversal occurrence probability &
0.25 \\
PSTOCH &
Stochastic Traversal occurrence probability &
0.25 \\
RAND5 &
Transaction root object random distribution &
Uniform \\
CLIENTN &
Number of clients &
1 \\
\hline
\end{tabular}
\vspace{5pt}
\caption{OCB workload parameters}
\label{tab-2}
\end{center}
\end{table}

        The metrics measured by OCB are basically the database response time (global, and per
transaction type), the number of accessed objects (still globally, and per transaction type), and
the number of I/O performed. We distinguish the I/Os necessary to execute the transactions,
and the clustering I/O overhead (I/Os needed to cluster the database).

\section{Validation Experiments}

        The Object Clustering Benchmark has been used to measure the performances of the
DSTC clustering technique, that is implemented in the Texas system, running on a Sun workstation.
The efficiency of DSTC had already been evaluated with another benchmark called
DSTC-CluB \cite{BULLAT96a}, that was derived from OO1. Hence, some comparisons are possible.

\subsection{Texas and the DSTC Clustering Technique}

        Texas is a persistent storage for C++, designed and developed at the University of Texas,
Austin \cite{SINGHAL92}. Texas is a virtual memory mapped system, and is hence very efficient.
Memory data formats are those of C++. Persistent objects are stored on disk using this same
format. Thus, when a disk page is loaded into memory, all the disk addresses towards referenced
objects are swizzled to virtual memory addresses, and vice versa, when a page is
trashed from the virtual memory.

        The {\em Dynamic, Statistical, and Tunable Clustering} technique has been developed at Blaise
Pascal University (Clermont-Ferrand~II) as a Ph.D. project \cite{BULLAT96b}. DSTC is heavily
based on the observation of database usage (basically, inter-object links crossings). It utilizes
run-time computed statistics to dynamically reorganize the database whenever necessary. The
DSTC strategy is subdivided into five phases.
\begin{enumerate}
\item {\em Observation Phase:} During a predefined Observation Period, data related to the transactions
execution is collected and stored in a transient Observation Matrix.
\item {\em Selection Phase:} Data stored in the Observation Matrix are sorted. Only significant statistics
are saved.
\item {\em Consolidation Phase:} The results of the Selection Phase are used to update the data gathered
during the previous Observation Periods, that are stored in a persistent Consolidated Matrix.
\item {\em Dynamic Cluster Reorganization :} The Consolidated Matrix statistics are used either to
build new Clustering Units, or to modify existing Clustering Units.
\item {\em Physical Clustering Organization:} Clustering Units are eventually applied to consider a
new object placement on disk. This phase is triggered when the system is idle.
\end{enumerate}

\subsection{Material Conditions}

        DSTC is integrated in a Texas prototype (version 0.2.1) running on a Sun SPARC/ELC
workstation. The operating system is SUN OS version 4.3.1. The available main memory is
8 Mb, plus 24 Mb of disk swap. The disk is set up with pages of 4 Kb. Texas and the additional
DSTC modules are compiled using the GNU C++ (version 2.4.5) compiler.

\subsection{Experiments and Results}

        These results are not the outcome of extensive tests performed on the Texas / DSTC couple.
They are rather significant experiments to demonstrate OCB's feasibility, validity, and
functionality.

\subsubsection{Object Base Creation Time:}

\begin{figure}[htb]
\epsfxsize=17cm
\vspace*{-0.8cm}
\centerline{\epsffile{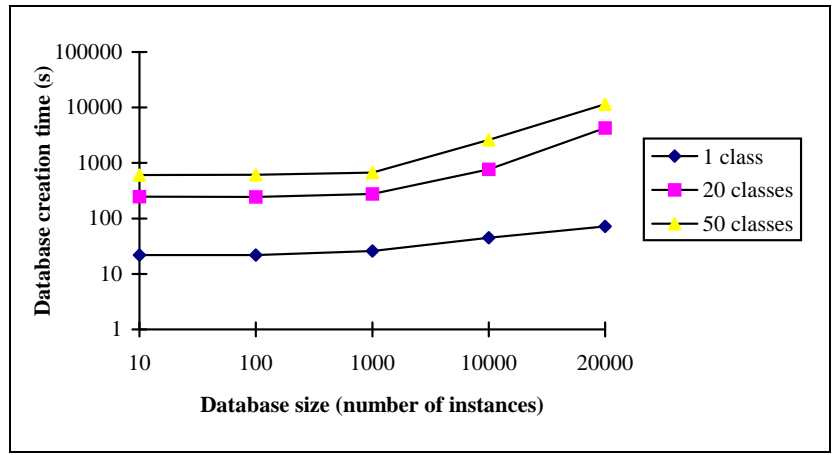}}
\vspace*{-16.5cm}
\caption{Database average creation time, function of the database size}
\label{fig-8}
\end{figure}

        The aim of this series of tests is to demonstrate the mere feasibility of OCB. Since we
recommend the use of a complex object base, we had to check if our specifications were possible
to achieve. Figure~\ref{fig-8} presents the database average generation time depending on the database
size, with a 1-class schema, a 20-class schema, and a 50-class schema. The time required
to create the object base remains indeed reasonable, even for the biggest OCB database
(about 15 Mb) used with Texas. The number of classes in the schema influences the database
generation time because the inheritance graph consistency is rendered more complex by a high
number of classes.

\subsubsection{I/O Cost:}

        We obtained a first validation for our benchmark by comparing the results achieved by
DSTC-CluB (the DSTC Clustering Benchmark) \cite{BULLAT96a} with those achieved by OCB.
First, we tuned OCB's database so that it approximates DSTC-CluB's (the OCB database parameters
values are provided by Table~\ref{tab-3}). The results, in terms of I/O cost, are sum up in
Table~\ref{tab-4}. Note that DSTC-CluB measures the number of transaction I/Os before, and after the
DSTC algorithm reorganizes the database.

\begin{table}[htb]
\begin{center}
\begin{tabular}{|c|c|c|}
\hline
{\bf Name} &
{\bf Parameter} &
{\bf Value} \\
\hline
NC &
Number of classes in the database &
2 \\
MAXNREF &
Maximum number of references, per class &
3 \\
BASESIZE &
Instances base size, per class &
50 bytes \\
NO &
Total number of objects &
20000 \\
NREFT &
Number of reference types &
3 \\
INFCLASS &
Inferior bound, set of referenced classes &
0 \\
SUPCLASS &
Superior bound, set of referenced classes &
NC \\
INFREF &
Inferior bound, set of referenced objects &
PartId - RefZone \\
SUPREF &
Superior bound, set of referenced objects &
PartId + RefZone \\
DIST1 &
Reference types random distribution &
Constant \\
DIST2 &
Class references random distribution &
Constant \\
DIST3 &
Objects in classes random distribution &
Constant \\
DIST4 &
Objects references random distribution &
Special \\
\hline
\end{tabular}
\vspace{5pt}
\caption{OCB database parameters in order to approximate DSTC-CluB's database}
\label{tab-3}
\end{center}
\end{table}

\begin{table}[htb]
\begin{center}
\begin{tabular}{|c|c|c|c|}
\hline
{\bf Benchmark} &
{\bf Number of I/Os} &
{\bf Number of I/Os} &
{\bf Gain Factor} \\
&
{\bf (before reclustering)} &
{\bf (after reclustering)} &
\\
\hline
DSTC-CluB &
66 &
5 &
13,2 \\
\hline
OCB &
61 &
7 &
8,71 \\
\hline
\end{tabular}
\vspace{5pt}
\caption{Texas/DSTC performance, measured with OCB, and DSTC-CluB}
\label{tab-4}
\end{center}
\end{table}

        Though the actual values are a little different (due to the size of objects, that is constant in
DSTC-CluB, and varies from class to class in OCB), the general behavior remains the same.
We have just showed that OCB can be tuned to mimic the behavior of another benchmark,
and thus illustrated its genericity.

        In a second phase, we benchmarked Texas and DSTC using OCB parameters default values
(Table~\ref{tab-2}). The results, presented in Table~\ref{tab-5}, show that OCB indicates a lesser performance
for DSTC than DSTC-CluB. This is because DSTC-CluB has only one type of transaction
(OO1's traversal) running on a semantically limited object base. Since OCB runs several types
of transactions on the database (thus more closely modeling a real transaction workload), the
access patterns, and the associated usage statistics, are much less stereotyped. Hence, clustering
with DSTC is demonstrated not to be as good as stated in \cite{BULLAT96a}. It though remains
an excellent choice, since it still improves Texas' performances by a 2.5 factor.

\begin{table}[htb]
\begin{center}
\begin{tabular}{|c|c|c|c|}
\hline
{\bf Benchmark} &
{\bf Number of I/Os} &
{\bf Number of I/Os} &
{\bf Gain Factor} \\
&
{\bf (before reclustering)} &
{\bf (after reclustering)} &
\\
\hline
OCB &
31 &
12 &
2,58 \\
\hline
\end{tabular}
\vspace{5pt}
\caption{Texas/DSTC performance, measured with OCB}
\label{tab-5}
\end{center}
\end{table}

\section{Conclusions and Future Research}

        OCB has been demonstrated to be valid to benchmark the performances of clustering algorithms
in OODBs. We have indeed proved that, properly customized, OCB could confirm
results obtained with the previous benchmark DSTC-CluB, and even provide more insight
about clustering policies performance than DSTC-CluB (and, {\em a fortiori}, OO1).

        OCB's main qualities are its richness, its flexibility, and its compactness. OCB indeed
offers an object base which complexity has never been achieved before in object-oriented
benchmarks. Furthermore, since this database, and likewise, the transactions running on it, are
wholly tunable through a set of comprehensive (but easily set) parameters, OCB can be used
to model any kind of object-oriented database application. Lastly, OCB's code is short, reasonably
easy to implement, and easily portable, even in non object-oriented environments like QNAP2.

        The perspectives opened by this study are numerous. First, we have mentioned that OCB
could be easily enhanced to become a fully generic object-oriented benchmark. Since OCB's
object base is already generic, this goal can be achieved by extending the transaction set so
that it includes a broader range of operations (namely operations we discarded in the first
place because they couldn't benefit from clustering). We think too, that existing benchmark
databases might be approximated with OCB's schema, tuned by the appropriate parameters.

        Another obvious future task is the actual exploitation of OCB: the benchmarking of several
different clustering techniques for the sake of performance comparison, and possibly each
time on different systems (OODB/OS couple). We also plan to integrate OCB into simulation
models, in order to benefit from the advantages of simulation (platform independence, {\em a priori}
modeling of non-implemented research prototypes, low cost).

        Eventually, when designing our benchmark, we essentially focused on performance.
However, though very important, performance is not the only factor to consider. Functionality
is also very significant \cite{BERRE91}. Hence, we plan to work in this direction, and add a
® qualitative ¯ element into OCB, a bit the way \cite{KEMPE95} operated for the CAD-oriented
OCAD benchmark. For instance, we could evaluate if a clustering heuristic's parameters are
easy to apprehend and set up, if the algorithm is easy to use, or transparent to the user, etc.

\pagebreak

\end{document}